\documentstyle[aps,prd,preprint,epsfig,tighten]{revtex}
\begin{document}
\draft
\title{Indications on neutrino oscillations parameters  \\
        from initial K2K  and current SK data}
\author{	G.L.\ Fogli, 
		E.\ Lisi, 
		and A.\ Marrone\\[4mm]
}
\address{Dipartimento di Fisica and Sezione INFN di Bari\\
             	Via Amendola 173, 70126 Bari, Italy \\ }

\maketitle
\begin{abstract}
We briefly discuss the impact of initial data from the KEK-to-Kamioka (K2K)
neutrino experiment on the $\nu_\mu\to\nu_\tau$  oscillation parameters
($m^2,\tan^2\psi$) currently indicated by the Super-Kamiokande (SK) atmospheric
neutrino experiment.  After showing the very good agreement between K2K and SK,
we combine the two separate pieces of information.  We find that the 99\% C.L.\
range for $m^2$ allowed by SK only,  $m^2\in[1.3,\,5.6]\times10^{-3}$ eV$^2$,
is reduced to  $[1.5,\,4.8]\times 10^{-3}$ eV$^2$ by including K2K data. By
halving the uncertainties of the K2K total rate  (with central value
unchanged),  the $m^2$ range would be ulteriorly reduced to
$[1.8,\,4.0]\times10^{-3}$ eV$^2$. Such information appears to be already
useful in planning (very) long baseline neutrino oscillation experiments.
\end{abstract}
\medskip

\section{Introduction}

The KEK-to-Kamioka (K2K) long baseline neutrino experiment \cite{K2K0}  is
designed to explore, with laboratory neutrinos, the mechanism of $\nu_\mu$
disappearance  indicated by the Super-Kamiokande (SK) \cite{Evid}, MACRO
\cite{MACR}, and  Soudan~2 \cite{SOUD} atmospheric neutrino experiments. Such
mechanism appears to be dominated by $\nu_\mu\to\nu_\tau$ oscillations
\cite{3atm,Subd,Revi,Stau} with mass-mixing parameters $(m^2,\tan^2\psi)$ close
to $(3\times 10^{-3}{\rm\ eV}^2,\,1)$. The pathlength $(L=250{\rm\ km})$ and
typical energies ($E\sim{\rm few\ GeV}$) of K2K $\nu_\mu$'s are well suited to
study $\nu_\mu$ disappearance effects for such parameter values \cite{K2K0}.

The recent data released by the K2K experiment \cite{Ni01,Ju01,Ha01}  (44
observed neutrino events {\em vs\/} 63.9 expected) are already inconsistent
with the no-oscillation hypothesis at about 97\% C.L.\
\cite{Ni01,Ju01,Ha01,Kwww}, and suggest disappearance of the muon neutrino
flavor in the path from KEK to Kamioka. The K2K collaboration is being
understandably  rather conservative on the oscillation explanation of the data,
pending more accurate estimates of the K2K neutrino spectrum and interaction
uncertainties by means of the near detector \cite{KMRD}. However, it is
tempting to use just the minimum amount of  spectrum-integrated data (i.e., the
total event rate, for which the error  estimate appears to be relatively
stable), in order to study the compatibility of K2K and SK data \cite{Li01}.

In this paper we  attempt to investigate such a compatibility issue, by
comparing predictions and data in both K2K and SK, within the simplest
(two-family) $\nu_\mu\to\nu_\tau$ oscillation framework.  We show  that the K2K
and the SK data are consistent, and that their combination starts to be useful
in constraining further the oscillation parameters. In particular, the 99\%
C.L.\ range for $m^2$ allowed by SK only,  $m^2\in[1.3,\,5.6]\times10^{-3}$
eV$^2$, is reduced to  $[1.5,\,4.8]\times 10^{-3}$ eV$^2$ by including K2K
data. Finally we show that, by halving the K2K total rate error,  the $m^2$
range would be ulteriorly reduced to $[1.8,\,4.0]\times10^{-3}$ eV$^2$.

\section{Analysis of K2K}

Analyses of K2K data have to consider the energy spectrum of parent neutrinos
at the far detector $S(E)$, i.e., the spectrum of neutrino that have produced a
detected interaction at the Kamioka site.  Such spectrum is, in general, given
by the product of the unoscillated neutrino energy spectrum $\Phi$ at the far
(SK) detector, times the interaction cross section $\sigma$, times the
detection efficiency $\varepsilon$ (all in differential form), integrated over
all the final state parameters (simbolically, $S=\int
\Phi\cdot\sigma\cdot\varepsilon$).  The spectrum $S(E)$ is important since, in
the presence of neutrino mass and mixing,  the oscillated event rate  can be
expressed as $r=\int dE\;S(E)\cdot P_{\mu\mu}(E)$ [to be compared with the
unoscillated rate, $r_0=\int dE\;S(E)$], where $P_{\mu\mu}$ is the $\nu_\mu$
survival probability.

At present, there is enough public information on $\Phi$ \cite{Ni01,Ju01,Ha01},
but not yet on the product $\sigma \cdot\varepsilon$ relevant for the far
detector specifications. Therefore, a direct and accurate reconstruction of the
spectrum $S(E)$ and of its uncertainties  is not possible outside the K2K
collaboration, at least at present. However,  an indirect, approximate
reconstruction can be made, by considering that: (a) the neutrino spectrum is
practically zero above 5 GeV \cite{Ni01}; and (b) according to the K2K
MonteCarlo (MC) simulation, for the current exposure ($3.85\times 10^{19}$
p.o.t.) one has $r_0=63.9$ events, while $r=41.5$, 27.4, and 23.1 events for
$m^2=3$, 5, and $7\times 10^{-3}$ eV$^2$, respectively, assuming maximal mixing
\cite{Ni01,Ju01,Ha01}.  We have then introduced an empirical functional  form
for the spectrum, $S(E)\propto x^\alpha(1-x)^\beta$ with  $x=E/(5 {\rm\
GeV})$,   which is found to reproduce well the above MC results for 
$(\alpha,\,\beta)=(1.50,\,3.34)$. In the following analysis we use such
interacted neutrino spectrum,  since it gives---by construction---results close
to the K2K MC simulation, after integration over $P_{\mu\mu}(E)$. A more
accurate (and less tentative) parametrization of $S(E)$ shall be possible 
after the experimental specifications and selection cuts (relevant to determine
$\sigma \cdot \varepsilon$ and its uncertainties) will be measured and
described in detail.

Concerning the uncertainties, the current systematic error on the total rate is
estimated to be 10\% by the K2K Collaboration  \cite{Ni01,Ju01,Ha01}. Spectral
bins may be affected by larger (and correlated) errors \cite{Ni01,Ju01,Ha01},
which, however, do not affect our spectrum-averaged analysis. Therefore, we
attach a $10\%$ fractional error to the  oscillated number of events
$r=r(m^2,\tan^2\psi)$, while the statistical (Poisson) error $\sigma_{\rm
stat}=\sqrt{44}$ is attached to the number of observed events,  $r_{\rm
exp}=44$. The $\chi^2$ statistics for K2K is then simply given by  $(r-r_{\rm
exp})^2/\sigma^2_{\rm tot}$, where  $\sigma^2_{\rm tot}=\sigma^2_{\rm
stat}+(0.1\,r)^2$. With such definition, the no oscillation hypothesis
($r=r_0$) is disfavored at 97\% C.L.\  ($\chi^2=4.7$ for $N_{\rm DF}=1$),
consistently with the current claim of the K2K Collaboration
\cite{Ni01,Ju01,Ha01,Kwww}.

Figure~1 shows the  number $r$ of K2K events for the current exposure ($3.85
\times 10^{19}$ p.o.t.), as a function of $m^2$, for maximal mixing
$(\tan^2\psi=1)$. The solid curve, corresponding to our calculation of 
$r(m^2)$, starts from $63.9$ events in the no-oscillation limit ($m^2\to 0$),
and tends to the asymptotic value of 63.9/2 in the fast oscillation limit
($m^2\to\infty$), after a few ``wiggles'' associated to the first oscillation
cycles. The curve is very close to the four K2K MC points used to benchmark our
calculation. The horizontal gray band represents the $\pm 1\sigma_{\rm tot}$
interval  for the total number of observed events in K2K. The allowed band
disfavors no oscillations at $2.2\sigma$, and the first deep oscillation
minimum at $2.7\sigma$. It is instead in very good agreement with the
oscillated predictions for $m^2\sim{\rm few}\times 10^{-3}$ eV$^2$, the best
fit being reached at $m^2=2.7\times 10^{-3}$ eV$^2$. Such a value is very close
to the one independently quoted by the SK Collaboration as the current best fit
to their atmospheric neutrino data  ($m^2=2.5\times 10^{-3}$ eV$^2$)
\cite{Tots}, under the same $\nu_\mu\to\nu_\tau$ oscillation hypothesis.%
\footnote{It is also very close to the best fit of upgoing muon data in MACRO,
$m^2=2.4\times 10^{-3}$ eV$^2$ \protect\cite{MACR}.}
Therefore, it makes sense to combine K2K and SK data, as we do in the following
section.

\section{K2K and SK: combination and  and implications}

Given the very good agreement between the K2K and SK allowed ranges for $m^2$
at maximal mixing, it is interesting to study their compatibility for
unconstrained $2\nu$ mixing. Concerning the SK atmospheric $2\nu$ analysis, we
make use of the results recently reported by us in \cite{Subd} (for 79.5 kTy SK
data), specialized to the simplest  scenario of pure $\nu_\mu\to\nu_\tau$
oscillations.%
\footnote{Therefore, in this work we use $N_{\rm DF}=2$ to draw  iso-$\Delta
\chi^2$ contours,  while we used $N_{\rm DF}=3$ for the more general $3\nu$ and
$4\nu$ cases considered in \protect\cite{Subd}.}

Figure~2 shows the regions allowed at 90\% and 99\% C.L.\  by SK, K2K, and
their combination, in the $2\nu$ mass-mixing plane $(m^2,\tan^2\psi)$. The
upper left panel corresponds to SK atmospheric data only, for which we find
two  degenerate best-fit points (stars) at  $m^2=3\times 10^{-3}$ eV$^2$ and at
octant-symmetric mixing values, $(\tan^2\psi)^{\pm 1}=0.76$, corresponding
to slightly nonmaximal oscillation amplitude, $\sin^22\psi=0.98$.%
\footnote{However, such small deviations from maximal mixing at best fit (also
found in  \protect\cite{Pena}) are not statistically significant at present.} 

The upper right panel in Fig.~2 shows the fit to K2K only.  The locus of
best-fit points is a continuous, octant-symmetric curve (not shown) passing
through $(m^2/{\rm eV}^2,\,\tan^2\psi)=(2.7\times 10^{-3},\,1)$. The K2K
constraints in the mass-mixing plane are weaker than those placed by SK,
especially on $\tan^2\psi$. Therefore, one cannot expect a significant
improvement on $\tan^2\psi$ limits from the SK+K2K combination. Concerning
$m^2$,  the no-oscillation limit $m^2\to0$ is still allowed at 99\% C.L., while
values around $m^2\simeq 8.5\times 10^{-3}$ eV$^2$ are excluded (for large
mixing).   Such values correspond to the first (deep) oscillation minimum in
Fig.~1, which is disfavored by the data at $2.7\sigma$, as previously noted.
Therefore, in the SK+K2K combination, we expect ``high'' values of $m^2$ to be
more disfavored than ``low'' values.

The lower left panel in Fig.~2 shows the combination of SK and K2K data. The
best-fit points are located at the same values of $\tan^2\psi$ as for SK alone
and, in general, the bounds on $\tan^2\psi$ are not significantly modified, as
expected. The best-fit value of $m^2$ is only slightly lowered ($m^2=2.9\times
10^{-3}$ eV$^2$) but, most importantly, the 90\% and 99\% C.L.\ ranges of $m^2$
are appreciably reduced both from below and (more strongly)  from above. Such
results show that the K2K experiment is already having a nonnegligible impact
in the determination of the neutrino squared mass  difference $m^2$ relevant
for the $\nu_\mu\to\nu_\tau$ channel and for the leading oscillations in (very)
long baseline experiments.

The lower right panel in Fig.~2 shows a prospective  SK+K2K combination, with
the same SK data and K2K data but with K2K total uncertainty hypothetically
reduced by a factor of two.  The $m^2$ range is significantly narrowed, while
the $\tan^2\psi$ range is still basically unchanged.  In summary, the various
panels of Fig.~2 demonstrate that K2K data are relevant for the determination
of $m^2$,  rather than of $\tan^2\psi$. Therefore, it makes sense to discuss in
more detail the results of Fig.~2 in terms of $m^2$ only, with $\tan^2\psi$
unconstrained.

Figure~3 shows the results of such an exercise, in terms of the function
$\chi^2(m^2)$. The 90\% and 99\% C.L.\ ranges for $m^2$ ($N_{\rm DF}=2$) are
explicitly shown for the SK and SK+K2K fits. Numerically, we find that the 99\%
C.L.\ range for $m^2$ allowed by SK only, $m^2\in[1.3,\,5.6]\times10^{-3}$
eV$^2$, is reduced to  $[1.5,\,4.8]\times 10^{-3}$ eV$^2$ by including K2K
data. By halving the uncertainties of the K2K total rate,  the $m^2$ range
would be ulteriorly reduced to $[1.8,\,4.0]\times10^{-3}$ eV$^2$
(99\% C.L.).

The reduction of the minimum values of $m^2$ in Fig.~3  (at a given C.L.) is
relevant for future (very) long baseline experiments. For instance, in the
OPERA experiment \cite{OPER} downstream the CERN-to-Gran Sasso neutrino beam,
the $\tau$ appearance rate is approximately proportional to $(m^2)^2$
\cite{OPER}; therefore, the increase of $m^2_{\min}$ from $1.3\times10^{-3}$
eV$^2$ (SK) to $1.5\times10^{-3}$  eV$^2$  (SK+K2K) at 99\% C.L.\ implies an
increase of the minimum expected $\tau$ event rate by a factor $\sim 1.3$ (at
the same C.L.); the increase would be as high as a factor $\sim 2$ for the case
with halved K2K errors ($m^2_{\min}=1.8\times 10^{-3}$ eV$^2$). The reduction
of the $m^2$ allowed range is also important to refine the energy or baseline
optimization in future neutrino factory experiments, where the current $m^2$
uncertainty plays an important role  (see, e.g., \cite{Opti} and references
therein). Finally, the increase of $m^2_{\min}$ strengthens the accuracy of
approximations based on the hierarchy of squared mass differences, which have
long been used in global solar+terrestrial neutrino analyses (see, e.g.,
\cite{Subd,Pena,Hier}).

A final remark is in order. Our current analysis is based on initial K2K data
and on an approximate description of the K2K detector specification. Therefore,
it cannot be a substitute of the (joint) official oscillation analysis that
will be performed by the K2K (SK+K2K) collaboration(s). However, we think that
our results, although necessarily approximate, are sufficiently indicative of
the K2K potential in improving our current knowledge of $\nu_\mu\to\nu_\tau$
oscillations. 

\section{Summary}

We have shown that, within the simplest (two-family) $\nu_\mu\to\nu_\tau$
oscillation scenario, the initial evidence for a $\nu_\mu$ flux suppression in
K2K is perfectly consistent with the  atmospheric $\nu_\mu$ flux anomaly. The
range of the neutrino squared mass difference indicated by the SK atmospheric
$\nu$ experiment is reduced by including the K2K data. As a consequence, K2K
starts to be important in narrowing the range of predictions for future (very)
long baseline experiment.

\acknowledgments

We thank T.\ Hasegawa and T.\ Maruyama for very useful information  about the
K2K experiment. E.L.\ thanks the organizers of the TAUP Conference, where
preliminary results of this work were presented, for kind hospitality. This
work was supported by the Italian Istituto Nazionale di Fisica Nucleare (INFN)
and Ministero dell'Istruzione, Universit\`a e Ricerca (MIUR) within the
``Astroparticle Physics'' project.


\newcommand{\InsertFigure}[2]{\newpage\begin{center}\mbox{%
\epsfig{bbllx=1.4truecm,bblly=1.3truecm,bburx=19.5truecm,bbury=26.5truecm,%
height=22truecm,figure=#1}}\end{center}\vspace*{-2.8truecm}%
\parbox[t]{\hsize}{\small\baselineskip=0.5truecm\hspace*{0.5truecm} #2}}
\InsertFigure{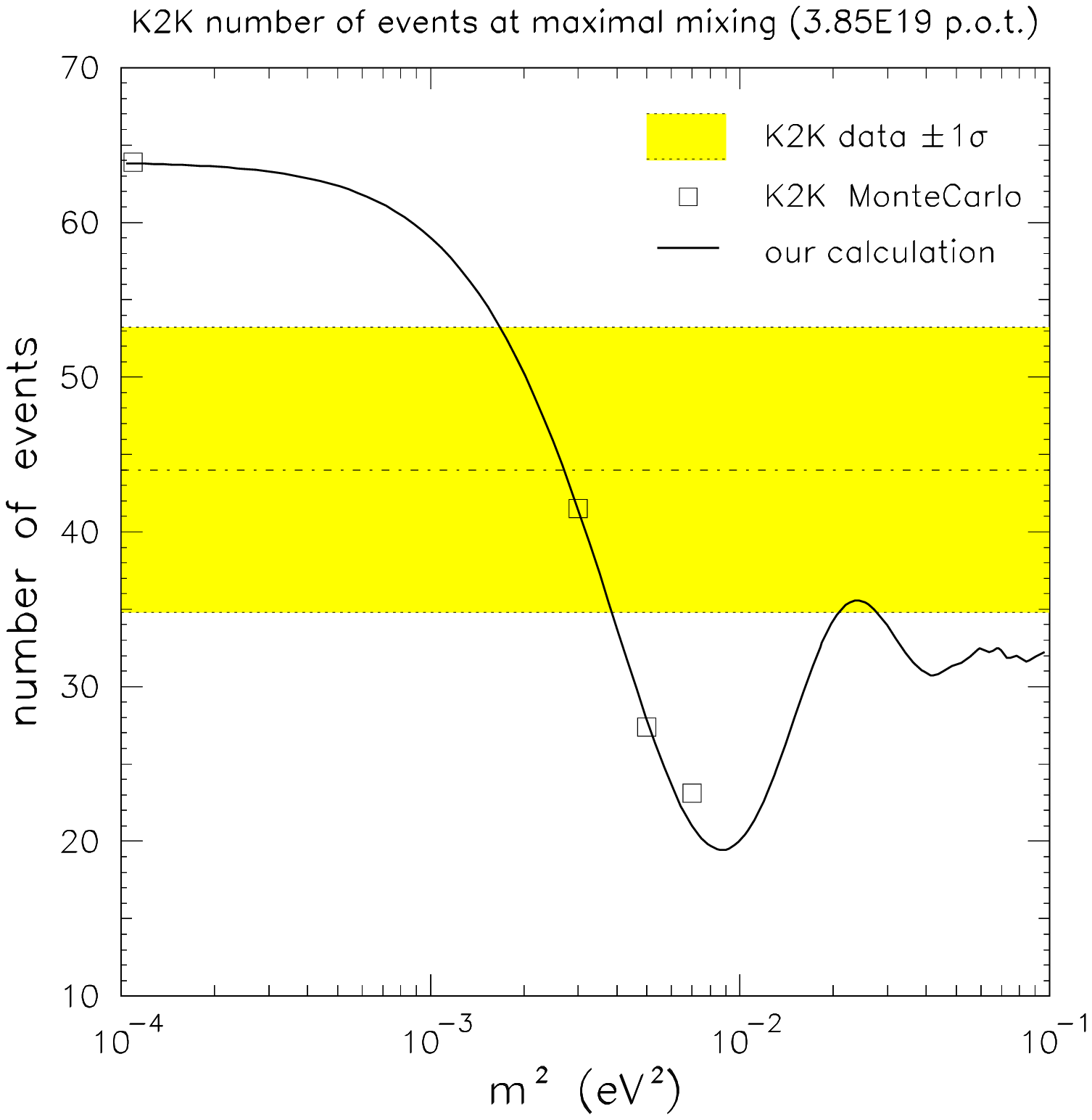}%
{FIG.~1. Number of events in K2K for the current exposure  ($3.85\times
10^{19}$ p.o.t.), as a function of $m^2$ (at $\tan^2\psi=1$). Our calculation
(solid curve) is benchmarked by the K2K MC simulation (square markers). The
horizontal gray band represents the current K2K data within one standard
deviation. }
\InsertFigure{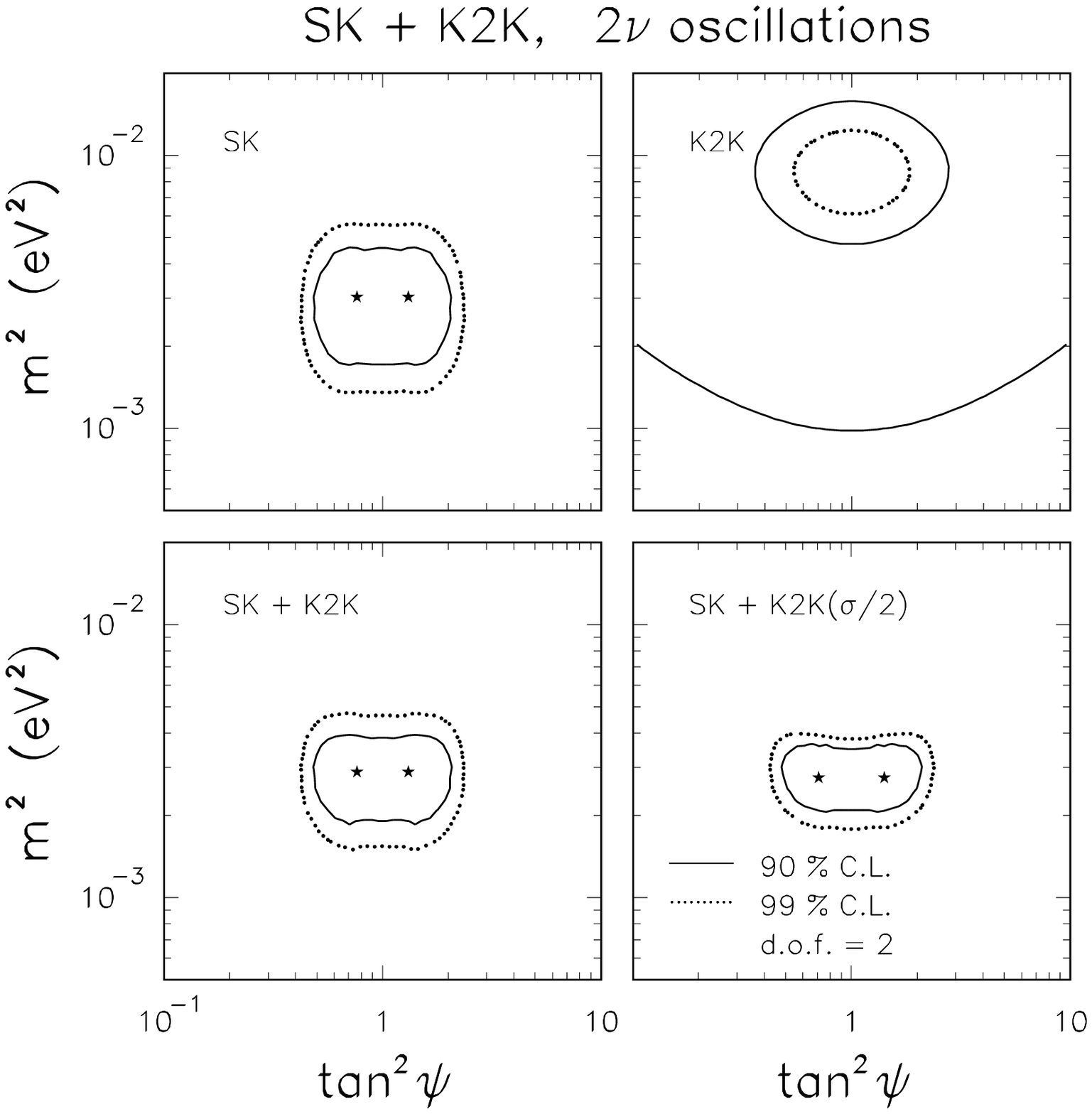}%
{FIG.~2. Separate two-flavor oscillation analyses of SK and K2K, together with
their combination (with eventual halving of K2K errors). See the text for
details.}
\InsertFigure{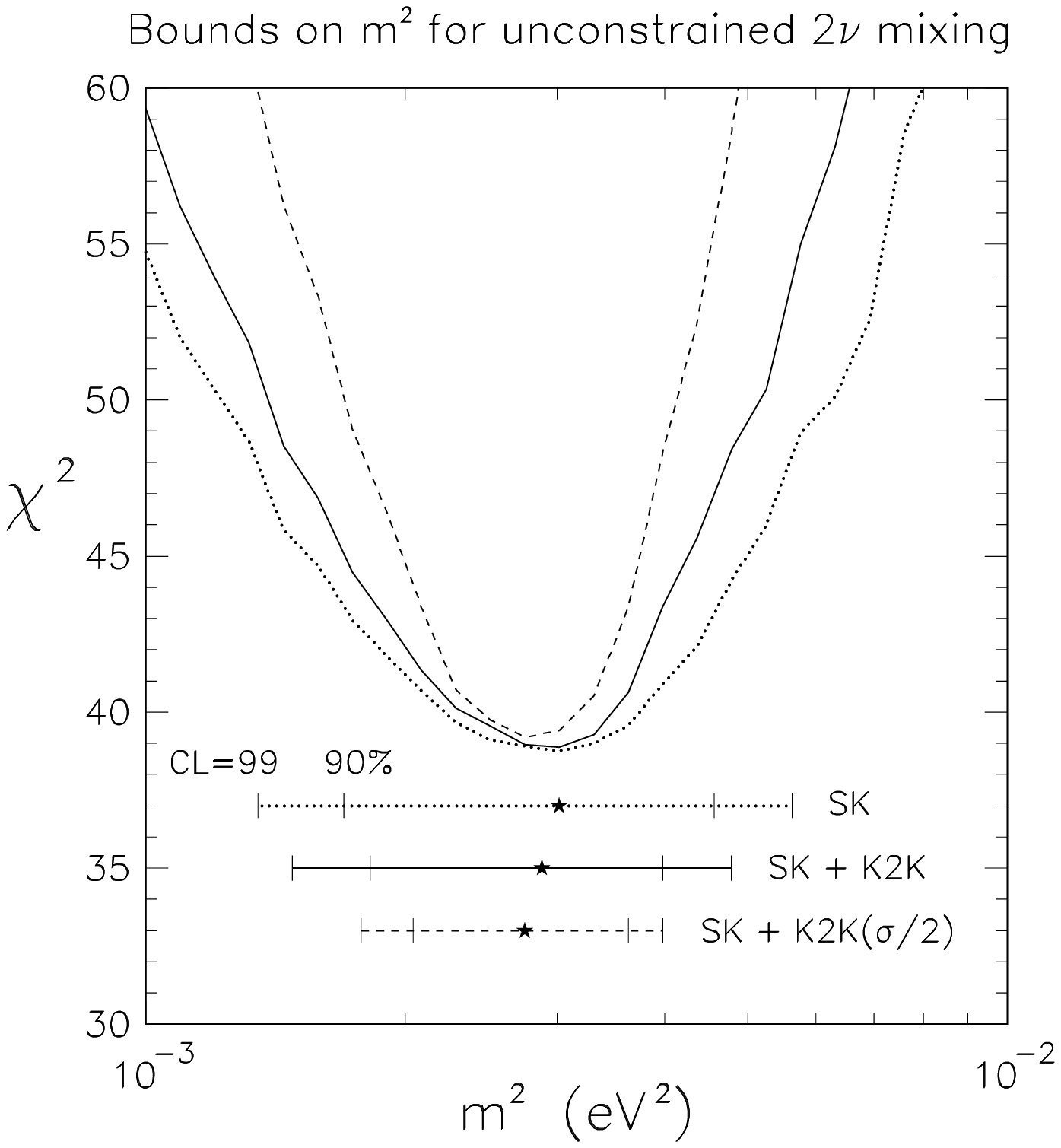}%
{FIG.~3. Bounds on $m^2$ at 90\% and 99\% C.L.\ ($N_{\rm DF}=2$)  from the
two-flavor $\chi^2$ analysis of the SK and K2K data.}

\eject

\begin{thebibliography}{99}

\bibitem{K2K0}	K2K Collaboration, S.H.\ Ahn {\em et al.},
		Phys.\ Lett.\ B {\bf 511}, 178 (2001).

\bibitem{Evid}	Super-Kamiokande Collaboration, Y.\ Fukuda {\em et al.},
		Phys.\ Rev.\ Lett.\ {\bf 81}, 1561 (1998). 	
		
\bibitem{MACR}	MACRO Collaboration, M.\ Ambrosio {\em et al.},
		Phys.\ Lett.\ B {\bf 517}, 59 (2001).

\bibitem{SOUD}	Soudan-2 Collaboration, W.W.M.\ Allsion {\em et al.},
		Phys.\ Lett.\ B {\bf 449}, 137 (1999).

\bibitem{3atm}	G.L.\ Fogli, E.\ Lisi, A.\ Marrone, and G.\ Scioscia,
		Phys.\ Rev.\ D {\bf 59}, 033001 (1999).

\bibitem{Subd}	G.L.\ Fogli, E.\ Lisi, and A.\ Marrone,
		hep-ph/0105139 (to appear in Phys.\ Rev.\ D).

\bibitem{Revi}	T.\ Kajita and Y.\ Totsuka,
		Rev.\ Mod.\ Phys.\ {\bf 73}, 85 (2001).

\bibitem{Stau}	Super-Kamiokande Collaboration, S.\ Fukuda {\em et al.},
		Phys.\ Rev.\ Lett.\ {\bf 85}, 3999 (2000). 	

\bibitem{Ni01}	K.\ Nishikawa,  
		in the Proceedings of the International  Europhysics Conference
		on High Energy Physics (Budapest,  Hungary, July 2001), to
		appear. Transparencies available at www.hep2001.elte.hu~.

\bibitem{Ju01}	C.K.\ Jung, 
		in the  Proceedings of the XX International Symposium on Lepton
		and  Photon Interactions at High Energies (Rome, Italy, July
		2001), to appear. Transparencies available at
		www.lp01.infn.it~.

\bibitem{Ha01}	T.\ Hasegawa, 
		in the   Proceedings of {\em TAUP 2001}, 7th International 
		Workshop on Topics in Astroparticle and Underground Physics
		(Laboratori Nazionali del Gran Sasso, Italy, Sept.\ 2001), to
		appear. Transparencies available at taup2001.lngs.infn.it~.

\bibitem{Kwww}	K2K experiment website, http://neutrino.kek.jp~.	

\bibitem{KMRD}	K2K Muon Range Detector Group, T.\ Ishii {\em et al.},
		hep-ex/0107041.

\bibitem{Li01}	E.\ Lisi 
		in {\em TAUP 2001\/} \protect\cite{Ha01}.
	
\bibitem{Tots}  Y.\ Totsuka 
		in {\em TAUP 2001\/} \protect\cite{Ha01}.

\bibitem{Pena}	M.C.\ Gonzalez-Garcia, 	M.\ Maltoni, and C.\ Pe{\~n}a-Garay,
		hep-ph/0108073.

\bibitem{OPER}	OPERA Collaboration, M.\ Guler {\em et al.},
		CERN Report  SPSC-2001-025, available at 
		operaweb.web.cern.ch~.

\bibitem{Opti}  M.\ Freund, P.\ Huber, and M.\ Lindner,
		hep-ph/0105071;
		Y.\ Wang, K.\ Whisnant, and B.\ Young,
		hep-ph/0109053.
	
\bibitem{Hier}	G.L.\ Fogli, E.\ Lisi, and D.\ Montanino,
		Phys.\ Rev.\ D {\bf 49}, 3626 (1994);
		Astropart.\ Phys.\ {\bf 4}, 177 (1995).

\end{thebibliography}
\end{document}